# SMASHNOVA MODEL

C. Sivaram

Indian Institute of Astrophysics, Bangalore 560034

The phenomenon of one celestial body smashing into another is quite common. This process on all scales can be very energetic. Recent example in the Solar system is that of the Comet Shoemaker-Levy, that slammed into Jupiter[1]. The fragments measuring 3 km across released 6 million megatons of energy. This is equivalent to one HIROSHIMA BOMB every second continuously for 10 YEARS

If a planet of Earth's mass collides with Jupiter [2], we can observe EUV-SOFT-XRAY flash for several hours and bright IR glow lasting for several thousand years. In dense stellar clusters, like globular cluster, star collisions are not uncommon. (LACK OF RG's near cores of GC's; LARGE C.S's for R.G's)

The origin of the BLUE STRAGGLERS in old stellar populations is due to merging of 2 or maybe 3 MS stars of 0.8 $M_\odot$. Perhaps about half the stars in central regions of some GC's underwent one or two collisions, over a period of $10^{10}$ years [3].
IN R136 cluster of TARANTULA nebula, there are more than $10^7$ stars in a region less than a parsec!

There are many examples of celestial bodies colliding. The collision of galaxies has been studied for long. The MILKY WAY and ANDROMEDA galaxies are approaching each other at ~300 km/sec. They are due for collision in another 3 billion years!

White Dwarf binary with less than 5 min period merges in a few thousand years. Neutron Star-White Dwarf binaries with periods, 11 min and 10.8 min are also observed which will undergo merger [4].

HEAD ON COLLISIONS

If a White Dwarf smashes into an MS star like the Sun, we need to know the signatures. The incoming velocity is $\geq$ 700 km/sec. The massive shock wave would compress and heat the sun.

The time taken for the 'SMASH UP' is about 5000s or that is about an hour.

The tidal energy released is given by $\dfrac{3GM_{WD}M_{\Theta}R_{\Theta}}{d^2}$ …(1)

This works out to be a few $\times 10^8$ deg.

Due to the impact the nuclear reactions become much faster. In about an hour, Sun would release thermonuclear energy of about $10^{49}$ ergs, as much energy as it would release in $2 \times 10^8$ years! On an average it will be about $3 \times 10^{45} ergs/\sec$. The instabilities would blow the sun apart in a few hours. The White Dwarf being much denser would continue on its way!

If a white dwarf (WD) impacts a red giant (RG), it would take about 2 months to penetrate the BLOATED RG. The RG would collapse, becomes another WD. If the white dwarfs merge, it can form a neutron star (NS). This will release about $10^{53}$ ergs of binding energy. NS impacting a RG or RSG can cause a SN outburst first followed by collapse of NS and the in-falling material into a black hole (BH) and hence leading to GRB. NS colliding with a WR star will result in SN followed by GRB, as the core collapses to a BH.

Black holes in a certain mass range can tidally disrupt a neutron star [5], leading to a $10^{53} erg$ GAMMA RAY BURST. In the case of WD and NS close binary, the WD can be tidally stretched or broken up when the separation is about $R_{WD}$.

CO white dwarf (close to $N_{ch}$) can detonate due to heating. Tidal energy of the order of $10^{50}$ ergs can heat WD to about $10^9$ deg. This is enough to detonate C. this can hence lead to an SN.

Enough material falls on NS at velocity $> 0.1c$. If the amount of matter falling in is $5 \times 10^{32} g$, it has $K.E \approx 10^{52} ergs$. On impact, gamma rays of nuclei energy $\geq$ 1 MeV is released with more than $10^{52} ergs$ in $\gamma$ - ray photons.

NS can be spun up and the flux squeezing can increase B. When NS slows down due to dipole radiation (Magnetar), in-falling matter can make it collapse to a BH releasing more than $10^{53}$ ergs, with the acceleration of particles due to the magnetic field.
Tidal stretching and heating can considerably increase thermonuclear (detonation) rates, especially carbon burning. This process is strongly dependent on the temperature.

The Zeldovich number is given by,

$$Z_e = \frac{T_{crit}}{T_b}\left(\frac{T_b - T_u}{T_b}\right) \qquad \ldots(2)$$

Where, $T_{crit}$ is the triggering temperature. $T_b$ and $T_u$ are the burnt and unburnt material temperatures respectively.
For a $z_e \approx 10$, we have peaked energy generation rates.

The flame speed is related to Markstein number, which is given by,

$$V_F = \left(\frac{\sigma_b}{(C_p)^2} \frac{aX^2_{Fu}}{T_{crit}} e^{\frac{-T_{crit}}{T_b}}\right) \qquad \ldots(3)$$

Where, $\sigma_b, c_p$ are the conduction and specific heat.

When convecting WD reaches density of $3 \times 10^9 g/cm^3$ and temperature of $T = 7 \times 10^8 K$, the ignition turns critical [6,7]. Nuclear energy generation time scale is comparable to convective turnover time $(10^2 s)$ and order of sound travel time $\approx 10 s$ (over a scale height of about 500 km).

Flame has a laminar speed and buoyancy [8]; the convection speed is of the order of $10^2 km/s$. The material is accelerated due to the off centre ignition. One solar mass can become convective.

$$^{12}C + {}^{12}C \rightarrow {}^{20}Ne + \alpha \ (OR) \ {}^{24}Mg.$$

$$\dot{\varepsilon}_{nuc} \approx 10^{25} \times 5 \times n({}^{12}C)\left(\frac{\rho}{10^9}\right) f_{es} r_{{}^{12}C} \ ergs/g/s \qquad \ldots(4)$$

Where,

$$r_{{}^{12}C} \approx 7 \times 10^{-16} \left(\frac{T}{7 \times 10^8}\right)^{30} \qquad \ldots(5)$$

$$f_{es} \approx 10^3 \left(\frac{\rho}{10^9}\right)^{2.4} \left(\frac{T}{7 \times 10^8}\right)^{-7} \qquad \ldots(6)$$

Therefore, we have,

$$\dot{\varepsilon}_{nuc} \approx 3 \times 10^{13} \left(\frac{T}{7 \times 10^8}\right)^{23} \left(\frac{\rho}{10^9}\right)^{3.3} ergs/g/s \qquad \ldots(7)$$

The nuclear specific energy due to the reaction is of the order of $5 \times 10^{17} n({}^{12}C) ergs/g$. Where, $n({}^{12}C)$ is the fraction of ${}^{12}C$.

The specific heat is due to ions, electrons and radiation. It is given by, [9]

$$C_P = \frac{3 N_A k_B}{2\overline{A}} + \frac{\pi^2 k^2}{\left(\rho_f/mc\right) m_e c^2} \rho N_A Y_e T + \frac{4 a T^3}{\rho} \qquad \ldots(8)$$

The temperature and density are given by,
$T = 7 \times 10^8 K, \rho = 2 \times 10^9 g/cc$.

Therefore the specific heat is given by,
$C_P \approx 10^{15} ergs/g/10^8 K$. For $T = 7 \times 10^8 K$,

$$\tau_{nuc} = \frac{C_P T}{24 \dot{\varepsilon}_{nuc}} \approx 10 \left(\frac{7 \times 10^8}{T}\right)^{22} \left(\frac{10^9}{\rho}\right)^{3.3} s \qquad \ldots(9)$$

The pressure is given by,

$$P = k\rho^{4/3} = 10^{27} \left(\frac{\rho}{2 \times 10^9}\right)^{4/3} dyne/cm^2 \qquad \ldots(10)$$

$\eta = r/a$ where, $a = \left(\frac{k}{\pi G \rho_0^{2/3}}\right)^{1/2} \approx 400 km$.

$$M(r) \approx \frac{4}{3}\pi r^3 \rho_0 \left(1 - \frac{3}{10}\eta^2\right) \qquad \ldots(11)$$

$$\rho(r) = \rho_0 \left(1 - \frac{1}{2}\eta^2\right) \text{ (Polytropic index)} \qquad \ldots(12)$$

$$\frac{dP}{dr} = -\frac{GM(r)}{r^2}\rho(r) \qquad \ldots(13)$$

$$\frac{dT}{dr} = \left(1 - \frac{1}{\Gamma}\right)\frac{T}{P}\frac{dP}{dr} \qquad \ldots(14)$$

$$T(r) = T_0 \left[1 - 10^{-2}\left(\frac{\rho}{2 \times 10^9}\right)^{2/3}\left(\frac{r}{10^7}\right)^2 \left(1 - \frac{1}{15}\eta^2\right)\right] \qquad \ldots(15)$$

$$L = 4\pi \int \dot{\varepsilon}_{nuc} \rho r^2 dr = 10^{45} ergs/s \left(\frac{\rho}{2 \times 10^9}\right)^{4.3} \left(\frac{T}{7 \times 10^8}\right)^{23} \int \left(\frac{r}{10^7}\right)^2 (1 - f(n)) \qquad \ldots(16)$$

The size of the region [10] is about 150 km with density $\rho = 2 \times 10^9 g/cc$.

For a recent review of the parameters see ref. [16].

The heat flux, $Q(r) = \frac{C(r)}{4\pi r^2} = \frac{\rho}{2} v_{conv} C_P \Delta T \qquad \ldots(17)$

$$v_{conv} = \left(\frac{2g\Delta\rho}{\rho}\right)^{1/2} t^{1/2} \approx \left(\frac{4\varepsilon r P L}{3 C_P T}\right)^{1/3} \qquad \ldots(18)$$

$$\Delta T \approx \left(\frac{2Q^2 T}{\rho^2 C_P^2 gl\Delta P}\right)^{1/3} \approx 50 km/s \left(\frac{7 \times 10^8}{T}\right)^{1/3}\left(\frac{L}{10^{45}}\right)^{1/3} \qquad \ldots(19)$$

The conduction $\sigma = \dfrac{4acT^3}{\rho k}$ ...(20)

$$\eta = \dfrac{10^9}{Z}\left(\dfrac{\rho}{10^9}\right) \approx 10^9 \, g/cm/s$$ ...(21)

$$\mathrm{Re} \approx \dfrac{\rho v_{conv} l}{\eta} = 10^{14}$$ ...(22)

$$\Pr \approx \dfrac{C_P \eta}{\sigma} \approx 10^{-2}$$ ...(23)

$$Ra \approx \dfrac{g l^3 \rho^2 C_P \Delta P \Delta T}{T \eta \sigma} \approx 10^{24}$$ ...(24)

$$Nu = \dfrac{Ql}{\sigma \Delta T} \approx 10^{12}$$ ...(25)

Kolgomorov length $= l\,\mathrm{Re}^{-3/4} \approx 10^{-3}\,cm$.

Effect due to tidal stretching (change of area) is quantified by the Karlovitz number, which is given by,

$$k_a = \dfrac{dA}{A} = \dfrac{A(x_u) - A(x_b)}{A(x_u)} \cong t_f \dfrac{1}{A}\dfrac{dA}{dx}$$ ...(26)

Where, $t_f$ is the flame thickness.

$$V_{F_t} = V_{F_l}^{\,o}\left(1 + (\dfrac{1}{L_e} - 1)\dfrac{Ze}{2}\,\mathrm{ka}\right)$$ ...(27)

Where, $L_e$ is the Lewis number.

$$V_{Fl} = V_{Fl}^{\,0}\left(1 + M_a k_a\right)$$ ...(28)

Change in the flame speed causes several flame instabilities.[11] The Landau – Darrieus instability gives us a mechanism for the accelerating burning rate of detonation in a White Dwarf.

For a density ratio $r = \dfrac{\rho_u}{\rho_b}$, the growth rate is given by,

$$\exp\left[\frac{rkV_{Fl}^0}{r+1}\left(\frac{r^2+r-1}{r}+kt_fM_a\left(kt_fMa+2\alpha-1+kt_fM_a\right)\right)\right] \qquad \text{...(29)}$$

Consider [12] the reaction $^{12}C+^{12}C$

$$\frac{dX_C}{dt}=-\frac{1}{12}X_C^2\rho N_A\lambda \qquad \text{...(30)}$$

$$N\lambda_A = 5\times 10^{26}\left(\frac{T_{9_1 crit}^{\frac{5}{6}}}{T_9^{\frac{3}{2}}}\right)\times \exp\left(\frac{-85}{T_{9_1 crit}^{\frac{1}{3}}}-2.2\times 10^{-3}T_9^3\right) \qquad \text{...(31)}$$

Where, $T_9=\dfrac{T}{10^9}$ and $X_C$ is the mass fraction of $^{12}C$

For each $^{24}Mg$ nucleus created, 13.9 MeV is released. The reactions would proceed further as:

$^{24}Mg(\alpha,\gamma)\ ^{28}Si,...$

$^{28}Si+^{28}Si \to Ni^{56}$

The NS can also shrink to a quark star (QS) by accretion of impacting white dwarf fragments. Accretion rate of the corresponding fall back material is given by, [13]

$$\dot{m}\approx 10^{29}\,g/s\left(\frac{\rho_{acc}}{10^7\,g/cc}\right)\left(\frac{R}{10km}\right)^{\frac{3}{2}}\left(\frac{M}{1.4M_\odot}\right)^{\frac{1}{2}} \qquad \text{...(32)}$$

Where, $\rho_{acc}$ is the density of accreted matter.

The rotational period of a newly formed NS (or QS) is of the order of 1-2ms. The magnetic field is of the order of $10^{13}-10^{15}\,G$.

Light cylinder Radius, $R_L=\dfrac{c}{\Omega}=95km\left(\dfrac{P}{2ms}\right)$  ...(33)

Magnetospheric radius $(R_{mag})$ is obtained from the relation,

Ram pressure of infalling matter $\cong$ magnetic field pressure

The slow down due to the magnetic dipole emission causes collapse of NS to BH. During this process jets are emitted along the rotational axes.

$$R_{mag} = \left(\frac{B^2 R^6}{2\dot{m}(2GM)^{\frac{1}{2}}}\right)^{\frac{2}{7}} = \left(\frac{B}{10^{15} G}\right)^{\frac{4}{7}} \left(\frac{10^{29} g/s}{\dot{m}}\right)^{\frac{2}{7}} \left(\frac{R}{10km}\right)^{\frac{12}{7}} \left(\frac{1.5 M_\Theta}{M}\right)^{\frac{1}{7}} 60km \qquad \ldots(34)$$

The corotation radius $R_{CO}$ is:

$$R_{CO} = 30km \left(\frac{M}{1.4 M_\Theta}\right)^{\frac{1}{3}} \left(\frac{P}{2ms}\right)^{\frac{2}{3}} \qquad \ldots(35)$$

Propeller regime: In falling material may be accelerated and hence carries away angular momentum ($10^{30}$ grams carrying away $10^{50}$ ergs in ten seconds). [14]

$$\dot{J}_{prop} = 10^{47} erg \left(\frac{\dot{m}}{10^{29} g/s}\right) \left(\frac{R_{mag}}{60}\right)^2 \left(\frac{2ms}{\rho_0}\right) \approx 10^{52} ergs \qquad \ldots(36)$$

The rotational energy carried away in jets is of the order of $10^{52}$ ergs, sufficient to power a short duration GRB.

We can classify the various scenarios as arising from the following impact possibilities or impact types:

| IMPACT POSSIBILITIES | RESULT |
| --- | --- |
| WD hits Red Giant (RSG) | WD + WD |
|  | WD + Disk + WD |
| NS hits RSG | NS or BH + Disk + WD |
|  | RSG → SN |
|  | NS |
| NS hits RG | NS or BH + Disk + WD |
|  | RSG → SN |
| NS hits NS | NS or BH + Disk |
| NS hits WD | NS or BH + Disk |
| Canonical → NS hits NS | NS or BH + Disk |
| WD hits WD | NS |
| WD hits MS | WD |
| MS hits MS | Depending mass<br><br>WD + WD<br>NS + NS<br>BH |
| MS hits RSG<br><br>MS hits RG<br><br>SG hits SG<br><br>SG hits RG<br><br>RG hits RG | WD + WD<br><br>NS + NS |

There are 28 different possibilities!

Let the masses of the colliding bodies are $M_1$ and $M_2$ such that, $M_1 >>> M_2$, with radii, $R_1$ and $R_2$. The glancing event problem $\approx \frac{R_2}{R_1}$

The energy is given by,

$$E \approx \frac{GM_1 M_2}{R_1} \qquad \ldots(37)$$

The velocity, $V \approx 600 km/s \left(\frac{M_\Theta}{R_\Theta}\right)^{\frac{1}{2}}$ ...(38)

$c_s \approx 10 - 50 km/s$.

The shock crossing time is given by, $\frac{R_2}{c_s} \approx t_s$

$$\frac{dE}{dt} \approx \frac{E}{t_s} \approx 10^{48} erg/\sec\left(\frac{M_\Theta}{r_\Theta}\right)\left(\frac{r_\Theta}{c_s}\right) \qquad \ldots(39)$$

$$T_{PK} = \left(\frac{\eta L_{Edd}}{4\pi R^2_L \sigma}\right)^{\frac{1}{4}} \qquad \ldots(40)$$

The mass of material ejected on impact is given by, [15]

$$\frac{m_{ej}}{m_0} \cong 0.1\left\{\frac{V_{imp}^2}{8}\left(\frac{\rho_{tar}}{m_{imp}}\right)^{\frac{1}{3}}\right\}^{0.48} \qquad \ldots(41)$$

$$\frac{m(>V_{esc})}{m_{imp}} \approx 0.1\left(\frac{\rho_{imp}}{\rho_{tar}}\right)^{0.2}\left(\frac{V}{V_{esc}}\right)^{1.2} \qquad \ldots(42)$$

The above equations follow from Impact theory. To obtain this the Hertz theory of collisions may be used.

**Reference:**

1. A. Molina and F. Moreno, Astr. Lett. Comm., $\underline{40}$, pp 1-8, 2000 (and reference therein)

   Chodas P et al, Coll. Of Comet Shoemaker-Levy with Jupiter, pp 1-30, ed. K. Noll et al, CUP 1996

2. astro ph/0308218

3. S. Zwart et al. AA, 348, 117-127, 1999

4. R. Tamm and H. Spruit, Ap.J $\underline{561}$, 329, 2001

   W. Chen and X. Li, A and A, $\underline{450}$, L1, 2006

5. Sivaram. C, I.A.U. Symp, 119, 1984

6. Kuhlen et al, ASP Conf. Ser., 293, 3D Stellar Evolution, ed. S. Turcotte et al (San Francisco ASI), 147, 2003

7. J. Niemeyer, Ap.J, 523, L57, 1999

8. A. Khokhlov et al, Ap.J, $\underline{478}$, 678, 1997

9. A. Alastuey and B. Jancovici, Ap.J, $\underline{236}$, 1034, 1978

   D. Porter and P. Woodward, Ap.J, $\underline{127}$, 159, 2000

   Kraichnan, Phys. Fluids, $\underline{5}$, 1374, 1962

10. F. Timmes, Ap.J, 528, 913, 2000

11. M. Reinetie et al, A and A, $\underline{391}$, 1167, 2002

12. Clayton D, Principles of Stellar Evolution and Neucleosynthesis, Chicago Univ. Press, 1983

    G. Caughlan and W. A. Fowler, At. Data Nuc. Data Tables, $\underline{40}$, 283, 1988

    J. Niemeyer et al, Ap.J, $\underline{471}$, 903, 1996

13. J. Price and S. Rosswog, Science, 312, 719, 2006

14. J. Lattimer et al, Science, $\underline{304}$, 536, 2004

15. A. Sills et al, Ap.J, $\underline{548}$, 323, 2001

    J. Hurley et al, astro-ph/0201217, 2002

16. W. Hillebrandt and J. Niemeyer, ARA and A, $\underline{38}$, 191, 2000